\documentclass{article}
\usepackage{natbib} 

     \usepackage[preprint]{neurips_2019}

\usepackage[utf8]{inputenc} 
\usepackage[T1]{fontenc}    
\usepackage{hyperref}       
\usepackage{url}            
\usepackage{booktabs}       
\usepackage{amsfonts}       
\usepackage{nicefrac}       
\usepackage{microtype}      
\usepackage{bm}
\usepackage{amsmath,amssymb,amsfonts}
\usepackage{algorithmic}
\usepackage{graphicx}
\usepackage{textcomp}
\usepackage{xcolor}
\usepackage{caption}
\usepackage{subcaption}

\title{Joint analysis of clinical risk factors and 4D cardiac motion for survival prediction using a hybrid deep learning network}

\author{No authors given}

\author{Shihao Jin  \\ Dept. of Mathematics\\ Imperial College London\\ \texttt{shihao.jin18@imperial.ac.uk} 
\AND Nicolò Savioli  \\ Dept. of Computing \\ Imperial College London\\ \texttt{nsavioli@imperial.ac.uk} \\
\and Antonio de Marvao \\ MRC London Institute of Medical Sciences \\ Imperial College London \\ \texttt{antonio.de-marvao@imperial.ac.uk}
\and Timothy JW Dawes \\ MRC London Institute of Medical Sciences \\ Imperial College London \\ \texttt{tim.dawes@imperial.ac.uk}
\and Axel Gandy \\ Dept. of Mathematics \\ Imperial College London \\ \texttt{a.gandy@imperial.ac.uk} \\
\and Daniel Rueckert  \\ Dept. of Computing \\ Imperial College Lodon\\ \texttt{d.rueckert@imperial.ac.uk}
\and Declan P O'Regan \\ MRC London Institute of Medical Sciences \\ Imperial College London \\ \texttt{declan.oregan@imperial.ac.uk}}

\begin{document}

\maketitle

\begin{abstract}
In this work, a novel approach is proposed for joint analysis of high dimensional time-resolved cardiac
motion features obtained from segmented cardiac MRI and low dimensional clinical risk factors to improve survival prediction in heart failure. Different methods are evaluated to find the optimal way to insert conventional  covariates into deep prediction networks. Correlation analysis between autoencoder latent codes and covariate features is used to examine how these predictors interact. We believe that similar approaches could also be used to introduce knowledge of genetic variants to such survival networks to improve outcome prediction by jointly analysing cardiac motion traits with inheritable risk factors.
\end{abstract}
    
\section{Introduction}

Heart failure affects twenty-six million people worldwide with increasing prevalence in an aging population \citep{savarese2017global}. Current tools for outcome prediction lack precision and are insensitive to the complex physiology of heart disease. Accurate risk stratification is important for ensuring individualised management and effective care to improve survival. Cardiac magnetic resonance imaging (CMR) is the gold standard for assessing biventricular function and recently computer vision techniques have been combined with supervised denoising autoencoders to learn motion features that are predictive of survival \citep{bello2019deep}.  To handle right-censored survival outcomes, this network uses a Cox partial likelihood loss function. However, this architecture takes only high dimensional motion data as inputs and does not take advantage of low dimensional clinical risk factors. We proposed three different approaches to introduce conventional covariates such as age, haemodynamic measurements and exercise capacity to Bello’s model.  We also adopted several statistical techniques, including canonical correlation analysis, to understand how new information interacts with prior knowledge.

\section{Method}

\begin{figure}[htp]
	\centering
	\begin{subfigure}[b]{0.3\textwidth}
		\centering
		\includegraphics[width=\textwidth]{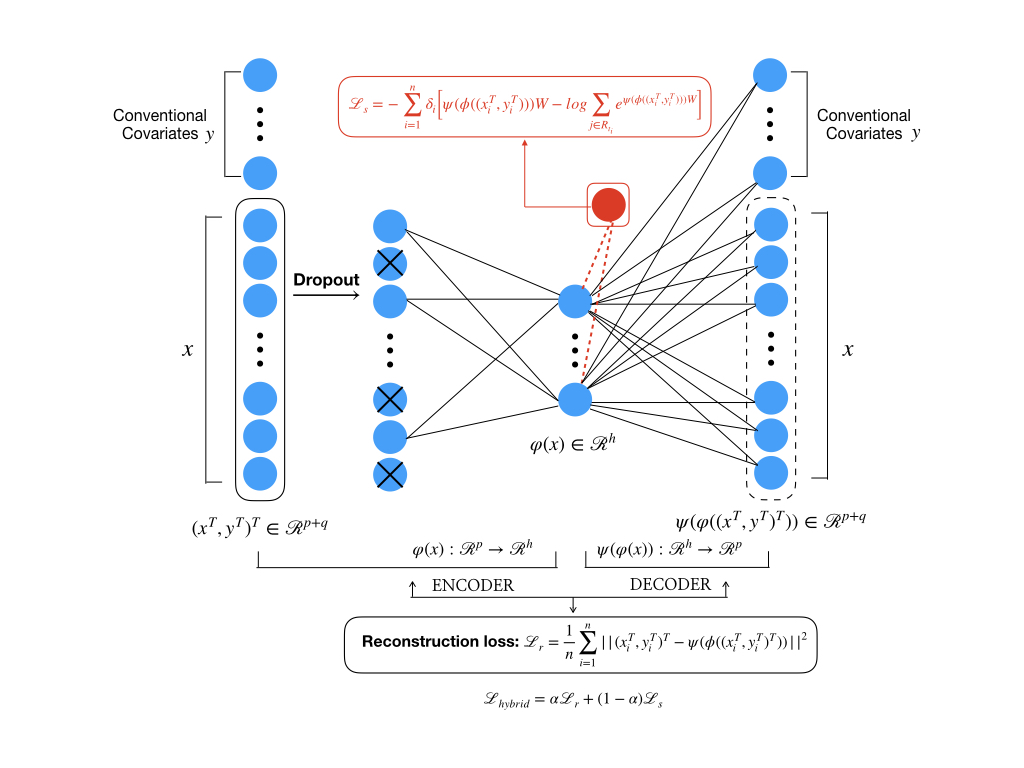}
		\caption{Model 1 - Extra information with input layer together with corruption and reconstruction process.}
		\label{model1}
	\end{subfigure}
	\hfill
	\begin{subfigure}[b]{0.3\textwidth}
		\centering
		\includegraphics[width=\textwidth]{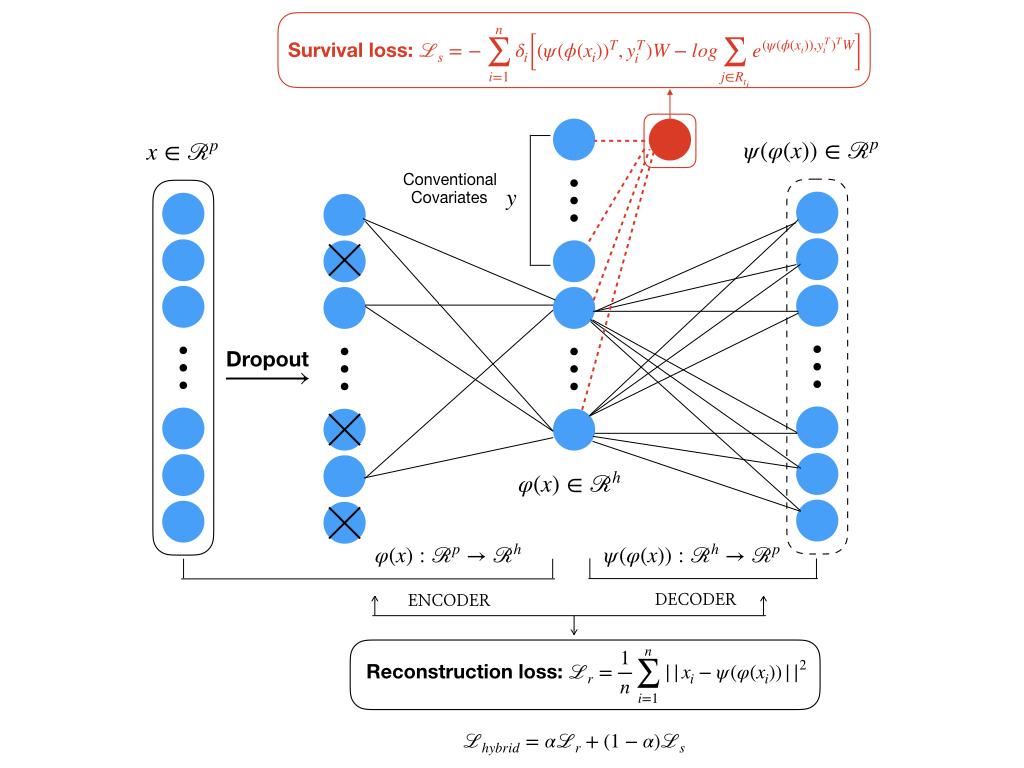}
		\caption{Model 2 - Extra information with latent code without corruption and reconstruction process.}
		\label{model2}
	\end{subfigure}
	\hfill
	\begin{subfigure}[b]{0.3\textwidth}
		\centering
		\includegraphics[width=\textwidth]{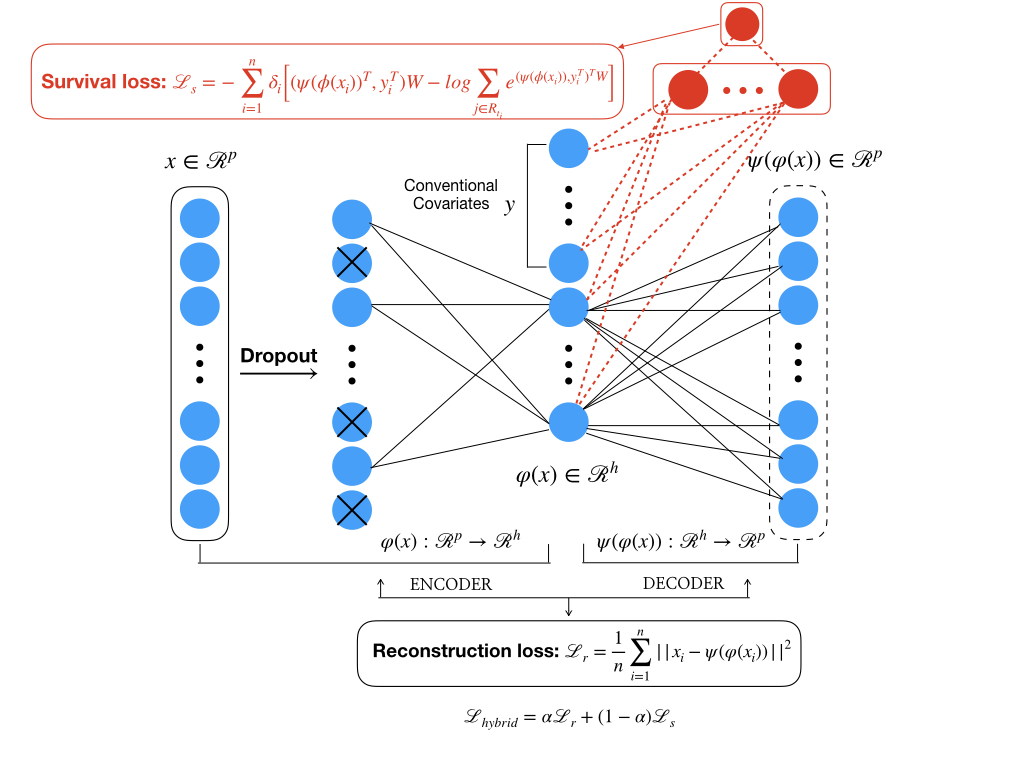}
		\caption{Model 3 - Extra information with latent code with a hidden layer between latent code and prediction layer.}
		\label{model3}
	\end{subfigure}
	\caption{The figure shows the three proposed models.}
	\label{model}
\end{figure}

In this work, three extensions of Bello's network (Model 0) are proposed to incorporate 
low dimensional clinical information. 
Model 1 (\ref{model1}) adds clinical data  to the input layer 
with corruption and reconstruction applied to both. Model 2 (\ref{model2}) aims 
to intensify the effect of clinical data by adding it with the latent code, 
without corruption or reconstruction. Model 3 (\ref{model3}) aims to model the nonlinear effect of clinical factors and interaction with the latent code using an additional hidden layer between the latent code and prediction layer.

\section{Experiments and Results}

A detailed description of the patient cohort and image analysis techniques can be found in \cite{bello2019deep}. 
We used eight new clinical factors in this work: age, sex, six-minute walk distance, 
functional class, 
mean pulmonary artery pressure and right ventricular (RV) 
end-diastolic volume, RV end-systolic volume, and 
RV ejection fraction. 
The models have several hyper-parameters. Indeed, in our experiments, Model 0, 1 and 2 have similar network structures (i.e they share the same set of hyper-parameters 
while Model 3 has a separate set of hyper-parameters). For model performance and validation 
we use Harrell's concordance index (C-index, \cite{harrell1982evaluating})
to measure the predictive accuracy based on bootstrap technique (due to small sample size).

\begin{table}[ht]
\centering
\renewcommand{\arraystretch}{1.5}
\begin{tabular}{ccc|ccc}
	\hline
	model & c-index & 95\% bootstrap CI & model & c-index & 95\% bootstrap CI\\
	\hline 
	Model 0  & 0.7979 & (0.7536, 0.8373)  \\ 
	Model 1  &  0.7978 & (0.7522, 0.8402) & Model 1 with noise & 0.8068 & (0.7644, 0.8457) \\ 
	Model 2  & 0.7998 & (0.7564, 0.8371) & Model 2 with noise & 0.8012 & (0.7548, 0.8407)\\ 
	Model 3  & 0.8276 & (0.7837, 0.8681) & Model 3 with noise & 0.8087 & (0.7718, 0.8433)\\ 
	\hline
\end{tabular} 
\vspace{0.1cm}
\caption{Model 0 represents the baseline model without clinical factors. 
Model 1, 2 and 3 (i.e described in methods section) are used for comparison.
Models with noise replacing clinical factors with 
standard gaussian noise and use 
the same hyper-parameters with the corresponding models where
95\% credible intervals are built by bootstrap.}
\label{tab1}
\end{table}

The results can be found in table \ref{tab1}. Numerically, Model 1 and 2 have no significant over-performance compared with Model 0, while Model 3 has better prediction ability than Model 0. Pair-wise two sample test based on bootstrap was also conducted for model comparison. The p-value for testing Model 3 being better than Model 0 is $0.0396$, significant at $0.05$ level. And the p-value for Model 1 v.s. 0 and Model 2 v.s. 0 are $0.5049$ and $0.4554$ respectively, revealing no improvement.

\begin{figure}[htp]
	\centering
		\includegraphics[width=1\textwidth]{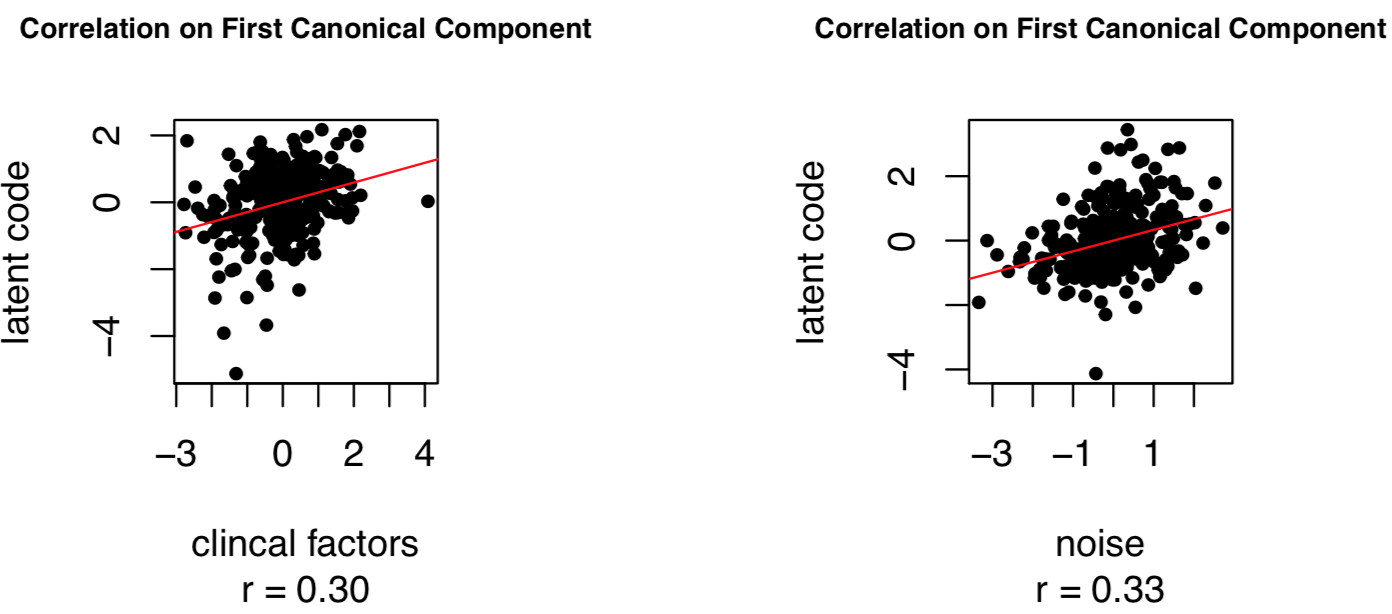}
		\label{model2cor1}
	\caption{Black points: individuals represented by first canonical component. Red line: fitted straight line by linear regression. r denotes the Pearson correlation between x-axis variable and y-axis variable.}
	\label{cor}
\end{figure}

Moreover, the preliminary correlation analysis shows that the Pearson correlation 
between all pairs of variables from latent code and clinical factors are below $0.2$, 
which is the same level of correlation between noise and latent code in Model 2.
Canonical Correlation Analysis (CCA, \cite{hotelling1992relations}) also shows that the latent 
code contains little information from clinical factors. Indeed, the overall linear correlation, 
mostly reflected by the first canonical component, between the latent code and clinical 
factors, are close to that of Gaussian noise (figure \ref{cor}).
Furthermore, to evaluate if clinical factors can linearly contribute to survival prediction, 
we also conducted linear regressions between the predicted risk and clinical factors; 
as well as between the predicted risk and latent code (table \ref{tab2}). 
Then, through the Model 2, we found that the latent code explained almost all of the predicted risk; 
whereas the clinical factors have a similar effect to noise for survival prediction. 
These findings suggest non-linear effect of the clinical risk factors modelled by Model 3.

\begin{table}[ht]
\centering
\renewcommand{\arraystretch}{1.5}
\begin{tabular}{ccc}
	\hline
	model & the predicted risk v.s. clinical factors & predicted risk v.s. latent code\\
	\hline 
	Model 2  & 0.02 & 0.84 \\ 
	Model 2 with noise & 0.03 & 0.95 \\
	\hline
\end{tabular} 
\vspace{0.1cm}
\caption{R-squares for regression models.}
\label{tab2}
\end{table}

\section{Discussion}

We found that combining routine clinical data with latent code from high dimensional cardiac motion can achieve improvements in survival prediction performance; as this allows for non-linear relationships and optimised the interaction between them. Such networks could be used in clinical practice for automated analysis of both cardiac imaging and routine clinical data for outcome prediction. Furthermore, this work suggests that other data, such as genetic variants and radiomic features, could be used in such survival networks to improve outcome prediction by jointly
analysing cardiac motion traits with inheritable risk factors.


\begin{thebibliography}{99} 

\bibitem[Savarese et al.(2017)]{savarese2017global} G.~Savarese and L.~H. Lund, Global public health burden of heart failure,
\emph{Card Fail Rev}, vol.~3, no.~1, p.~7, 2017.

\bibitem[Bello et~al.(2019)Bello, Dawes, Duan, Biffi, de~Marvao, Howard, Gibbs,
  Wilkins, Cook, Rueckert, O'Regan.]{bello2019deep}
G.~A. Bello, T.~J. Dawes, J.~Duan, C.~Biffi, A.~de~Marvao, L.~S. Howard,
  J.~S.~R. Gibbs, M.~R. Wilkins, S.~A. Cook, D.~Rueckert, D.~P. O'Regan.
\newblock Deep-learning cardiac motion analysis for human survival prediction.
\newblock \emph{Nat Mach Intell}, 1\penalty0 (2):\penalty0 95,
  2019.

\bibitem[Harrell et~al.(1982)Harrell, Califf, Pryor, Lee, and
 Rosati]{harrell1982evaluating}
F.~E. Harrell, R.~M. Califf, D.~B. Pryor, K.~L. Lee, and R.~A. Rosati.
\newblock Evaluating the yield of medical tests.
\newblock \emph{JAMA}, 247\penalty0 (18):\penalty0 2543--2546, 1982.

\bibitem[Hotelling(1992)]{hotelling1992relations}
H.~Hotelling.
\newblock Relations between two sets of variates.
\newblock In \emph{Breakthroughs in Statistics}, pages 162--190. Springer,
  1992.
  
  
    
 
\end{thebibliography}
\end{document}